\documentstyle[12pt]{article}

\begin{document}

\baselineskip=20pt

\renewcommand{\theequation}{\arabic{section}.\arabic{equation}}
\newfont{\elevenmib}{cmmib10 scaled\magstep2}

\newcommand{\YUKAWAmark}{\elevenmib
            Yukawa\hskip1mm Institute\hskip1mm Kyoto}

\begin{titlepage}

\begin{flushleft}
     \YUKAWAmark	 
\end{flushleft} \vspace{-40pt}
\begin{flushright}
     quant-ph/9611003\\February 1996
\end{flushright}

\vspace{0.5cm}

\begin{center}
                        
{\Large \bf    Generally Deformed Oscillator, Isospectral Oscillator}\\
\vspace{8pt}
{\Large \bf    System and Hermitian Phase Operator }

\vspace{1.2cm}

{\large \bf    Hong-Chen Fu\footnote{JSPS Fellow. On leave of absence 
                           from Institute of Theoretical Physics, 
                           Northeast Normal University, Changchun
                           130024, P.R.China.
                           E-mail: hcfu@yukawa.kyoto-u.ac.jp}  
               and 
               Ryu Sasaki\footnote{Supported partially by the 
                           grant-in-aid for Scientific Research, 
                           Priority Area 231 ``Infinite Analysis'' 
                           and General Research (C) in 
                           Physics, Japan Ministry of Education.}}

\vspace{0.5cm}

{\it           Yukawa Institute for Theoretical Physics, Kyoto 
               University\\ Kyoto 606-01, Japan }

\end{center}

\vspace{1cm}

\begin{abstract}
The generally deformed oscillator (GDO) and its 
multiphoton realization as well as the coherent and squeezed vacuum 
states are studied. We discuss, in particular, the 
GDO depending on a complex parameter 
$q$ (therefore we call it $q$-GDO) together with the finite dimensional 
cyclic representations. As a realistic physical 
system of GDO the isospectral oscillator system is studied 
and it is found that its coherent and squeezed vacuum states
are closely related to those of the oscillator. 
It is pointed out that starting from the $q$-GDO with $q$ root of unity
one can define the hermitian 
phase operators in quantum optics consistently and algebraically. 
The new creation and
annihilation operators of the Pegg-Barnett type phase operator theory 
are defined by using the cyclic representations and  these
operators  degenerate to those of the ordinary
oscillator in the classical limit $q \to 1$.

\end{abstract}

\vspace{0.9cm}

\begin{center}
     {\large  \sf Journal of Physics A: Mathematical and General}\\
     {\small \sf         29 (1996) 4049}
\end{center}

\end{titlepage}
                     

\newcommand{\p}{\mbox{e}^{i\hat{\Phi}_{\theta}}}
\newcommand{\rr}{\rangle}
\newcommand{\rref}[1]{(\ref{#1})}
\newcommand{\REF}[2]{(\ref{#1},\ref{#2})}
\newcommand{\RRR}[2]{(\ref{#1}-\ref{#2})}
\newcommand{\vv}{|\!|}

\section{Introduction}

Deformation of Lie algebras has been finding applications in various 
branches of physics. The $q$-deformed Lie algebras, or the quantum
algebras, play an important role in quantum integrable models
and quantum inverse scattering method \cite{defi,jimb}. 
The generally deformed oscillator (GDO)  first appeared in 
Heisenberg's theory of nonlinear spinor dynamics \cite{heis}.
In the literature, many more deformed oscillators can be found
\cite{arik,kury,bmsf,baye}  and a unification scheme for them has been 
suggested (see \cite{bon1} and references therein).  
Many physical systems are found to enjoy the GDO symmetry (for a list
see \cite{bon1,bon2,kara}). In this paper we shall pay attention 
to several types of GDO, namely the multiphoton realization of GDO 
and two new GDO systems: the isospectral oscillator system (ISOS) 
\cite{miel,nie1} and $q$-deformed GDO which is a
subalgebra of GDO. Based on the $q$-deformed GDO having  
finite dimensional {\it cyclic}
representations we construct the hermitian phase operator in
quantum optics algebraically.

In Sec.2 we first review the GDO and study its
multiphoton realization and the coherent and squeezed vacuum states. 
These states are expressed in terms of an exponential displacement 
operator acting on the vacuum state. 
We know that the GDO can be realized in 
terms of the usual single photon operator \cite{bon1,sha1} 
(multiphoton realization 
of some Lie algebras and the $q$-oscillator can be found in 
\cite{lie} and \cite{chan,katr,atak}, respectively, and the
single photon realization of $q$-oscillator has been
extensively studied \cite{many}), and that the exponential
coherent states of the GDO are already obtained in a different way
\cite{sha1}.
A new notion of `spontaneously broken' multiphoton realization
of $q$-oscillators is introduced here.

In Sec.3 we add a new member to the GDO family,
 the isospectral oscillator system (ISOS) \cite{miel}.
As suggested by the name, it
has the same spectrum as the oscillator. It can be formulated 
in the framework of supersymmetric quantum mechanics and factorization 
method \cite{nie1}. Its coherent states are  studied in \cite{fern}. 
We show that the creation and annihilation operators and the Hamiltonian
generate a GDO. We also study its coherent and squeezed
vacuum states and find that these states are closely related to the
density-dependent annihilation operator coherent states and to the
squeezed vacuum of the oscillator, respectively.

Section 4 is devoted to the connection between the hermitian
phase operator in quantum optics and the $q$-GDO with cyclic 
representations. The proper quantization of the phase angle of 
an oscillator was first considered by Dirac in 1927 \cite{dira}. 
For history and some review papers of the phase operator, 
see \cite{phsc}. However, because of the fact that the creation 
and annihilation operators of the oscillator do not admit a naive 
polar decomposition, ie. a product of  a
unitary  times a positive semi-definite hermitian operator, 
the problem kept unsolved for a
long time. Susskind and Glogower considered a weaker exponential phase
operator which is one-side unitary \cite{suss}, namely only one of 
the two relations $UU^\dagger=1$ or $U^\dagger U=1$ holds for the 
exponential phase operator $U=e^{i\phi}$. 
Recently it was  realized 
that the hermitian phase operator could be  defined in an
($S+1$)-dimensional Hilbert space  and that the expectation values of
physical quantities would tend 
to those of the oscillator in the limit $S \rightarrow \infty$ 
\cite{pegg,ell1}. Pegg and Barnett
considered  a {\it truncated oscillator} defined in the $(S+1)$-dimensional
space and presented the hermitian phase operator (PB phase operator)
\cite{pegg}. The truncated oscillator has one disadvantage from the 
symmetry point of view:
its operators do not form a closed algebra.
Moreover, the truncated oscillator is not the only way to realize 
the PB phase operator. For example, Ellinas revealed the relevance of the
PB phase operator to  the naive $q$-oscillator \cite{bmsf} with $q$ 
root of unity \cite{ell2}.
In his approach, however, the the naive $q$-oscillator 
provides only a finite dimensional space  and the hermiticity of the
phase operator is not automatically ensured because 
of the  use of  a {\it regular representation} (see Sec.4).
In order to  
connect GDO with the hermitian phase operator, we introduce a 
new GDO, which is a subalgebra of GDO depending on a complex 
parameter $q$ (therefor we call it $q$-GDO). It  has  finite 
dimensional cyclic representations when $q$ is a root of unity.  
We here
advance the problem in two points: (1) the $q$-GDO with $q$ root of
unity is particularly suited for  {\it algebraic realization} of 
 the PB phase operator theory; and (2) 
the cyclic representation of $q$-GDO (and therefore the
`$q$-oscillator') ensures automatically the {\it hermiticity} of the phase 
operator.

\section{Generally Deformed Oscillator }
\setcounter{equation}{0}

In this section we first review the GDO, then
 the coherent and the squeezed vacuum states of GDO are 
 presented explicitly in terms of an ordinary exponential operator.

\subsection{GDO and its multiphoton realization}

The GDO is an associative algebra {${\cal B}$ }  over the complex
number field C with generators
$ A^{\dagger}, A, \cal N$ and the unit 1 satisfying
\begin{equation} 
     \left[{\cal N}, A^{\dagger} \right]=A^{\dagger},\ \ \ 
     [{\cal N}, A]=-A, \ \ \  
     A A^{\dagger}=F({\cal N}+1),\ \ \ A^{\dagger} A = F({\cal N}),  
     \label{algebra}
\end{equation} 
where the hermitian non-negative function $F$ is called the 
{\it structure} function. 
It should satisfy the condition
\begin{equation} 
F(0)=0 \label{Fockcond}
\end{equation} 
in order to have the Fock representation.

The algebra (\ref{algebra}) can be realized in terms of the usual 
single photon operator \cite{bon1,sha1} (multiphoton realization 
of some Lie algebras and the $q$-oscillator can be found in 
\cite{lie} and \cite{chan,katr,atak}, respectively, and the
single photon realization of $q$-oscillator has been extensively 
studied \cite{many}). Here we would like to present the general 
multiphoton realization of GDO. For this purpose, we consider 
the multiphoton lowering operator
\begin{equation} 
       A= f(N)a^m,
\end{equation}  
where $a$ and $a^\dagger$ are the annihilation and creation
operators of the photon satisfying $\left[a, a^{\dagger}\right]=1$, 
$N=a^{\dagger}a$,  and $m$ is a positive integer. As usual the Fock 
states of the oscillator $a$ and $a^{\dagger}$ are denoted by 
$|n\rr$, $n=0,1,\ldots$; $a|0\rr=0$, $a|n\rr=\sqrt{n}|n-1\rr$,
$a^{\dagger}|n\rr=\sqrt{n+1}|n+1\rr$. The function $f(N)$ specifies 
the intensity dependent coupling, which is in general complex and 
we assume that $f(x)$ does not have zeros at non-negative integer 
values of $x$. By using $aa^{\dagger}=N+1$,
$a^2(a^\dagger)^2=(N+1)(N+2)$, etc, we obtain
\begin{eqnarray} 
    A A^{\dagger} &=& (N+1)(N+2)\cdots (N+m) f(N)f^*(N), \label{aada} \\
    A^{\dagger} A &=& (N-m+1)(N-m+2)\cdots N f(N-m) f^*(N-m).\label{aaad}
\end{eqnarray} 
It is obvious that we only need to restrict our discussion to the sector
$S_i$ ($i=0,1,\cdots,m-1$) spanned by the Fock states $|nm+i\rr$ ($n$ 
non-negative integers). Introducing the 
{\it multiphoton number operator} ${\cal N}_i \equiv {\cal N}$
($i=0,1,\cdots,m-1$) on the sector $S_i$
\begin{equation} 
      {\cal N} = \frac{1}{m}\left(N-i\right),\ \ \ i=0,1,\cdots,m-1,
\end{equation} 
and 
$F({\cal N}+1)\equiv (m{\cal N}+1+i)\cdots
(m{\cal N}+m+i)f(m{\cal N}+i)f^*(m{\cal N}+i)$, 
we can recast the
system (\ref{aada}),\thinspace (\ref{aaad})  in the following form
\begin{equation} 
     AA^{\dagger} = F({\cal N}+1),\ \ \  A^{\dagger} A=F({\cal N}), \ \ \ 
     \left[{\cal N}, A^{\dagger}\right]=A^{\dagger},\ \ \ 
     \left[{\cal N}, A\right]=-A,   \label{re}
\end{equation} 
which we call {\em intensity-dependent $m$-photon realization}
of {${\cal B}$ }.  Note that the r.h.s.
of \rref{aaad} vanishes on the Fock states $|n\rr$ for $0\leq n \leq m-1$, 
which implies $F(0)=0$ in each sector.

It is very tempting to apply the idea of  the system \rref{re} 
to the multiphoton ($m$-photon) realization of the $q$-deformed
oscillator. Let us choose  (here $q$ is a {\it real} deformation 
parameter)
\begin{equation} 
   f(N)\equiv \left\{ \frac{1}{(N+1)\cdots (N+m)}\left[
              \frac{N}{m}+1\right]\right\}^{\frac{1}{2}}\label{fqdef},
\end{equation} 
where $[x]\equiv (q^x-q^{-x})/(q-q^{-1})$,
and define
\begin{equation} 
      b_q \equiv A=f(N)a^m,\ \ \  
      b_q^{\dagger} \equiv A^{\dagger}=(a^{\dagger})^mf^*(N), \ \ \
      N_q \equiv {\cal N}+\frac{i}{m}. \label{q}
\end{equation} 
Then by using \rref{aada}\thinspace and \rref{aaad}\thinspace we would obtain 
{\em formally} the following  relations 
\begin{eqnarray} 
&&   b_qb_q^{\dagger} = [N_q+1], \ \ \
     \left[N_q, b_q^{\dagger}\right]= b_q^{\dagger}, \ \ \ 
     [N_q, b_q]=-b_q. \label{Nbb} \\
&&   b_q^{\dagger}b_q ={(N-m+1)(N-m+2)\cdots N\over{(N-m+1)(N-m+2)
     \cdots  N}}[N_q]= [N_q]. \label{bbdag}
\end{eqnarray} 
Eqs.\rref{Nbb},\rref{bbdag} are in fact a multiphoton
realization of the $q$-oscillator, 
\begin{equation} 
   b_qb_q^{\dagger}-qb_q^{\dagger}b_q=q^{-N_q}. \label{qboson}
\end{equation} 
It should be remarked that the eigenvalues of $N_q$ are not 
integers except for the $i=0$ sector.

By close inspection, however, one finds that the relation \rref{bbdag}
is not true in the ``vacuum'' of each sector $S_i$ ($i\ge1$ and 
$m>1$) (for the $m=1$ case see \cite{many}, the $S_0$ sector is 
discussed in \cite{chan}). Obviously the ``vacuum'' of the $i$-th 
sector $\vv 0\rr=|i\rr$  vanishes when applied by $b_q$,
\begin{equation} 
   b_q\vv 0\rr=f(N)a^m|i\rr=0, \quad i=0,1,\ldots,m-1. \label{bkill}
\end{equation} 
On the other hand, as remarked above, $[N_q]\vv 0\rr=[{i\over m}]
\vv 0\rr$ is {\it 
non-vanishing\/} for $i\ge1$. This apparent inconsistency is caused by
$0/0=1$ in \rref{bbdag}, since $N-i$ in the numerator and denominator vanish
on $\vv 0\rr=|i\rr$. To sum up, the relations 
(\ref{bbdag})and (\ref{qboson}) are broken only by the ``vacuum'' 
expectation value and all the other relations are correct. It would be 
very interesting if one could find physical applications of the 
`{\it spontaneously broken\/}' multiphoton realization of $q$-oscillator.

If we introduce the intensity and {\it sector-dependent} multiphoton 
coupling then we can obtain the $q$-oscillator in each sector.
Namely, if we define 
\[
   a_q=\sqrt{[{\cal N}+1]\over{(N+1)\cdots(N+m)}}\thinspace a^m 
\]
in each sector, then it is easy to see that
$a_qa_q^{\dagger}=[{\cal N}+1]$ and $a_q^{\dagger} a_q=[{\cal N}]$ 
are satisfied as 
operator equations. This result has been reported in \cite{katr} but they 
considered this realization only in the sector $S_0$.

\subsection{Coherent and squeezed vacuum states}

In this subsection we shall study the ladder-operator coherent and
squeezed vacuum states of the GDO. To this end we define a 
convenient orthonormal basis for $S_i$
\begin{equation} 
     \vv n\rr=\frac{1}{\sqrt{[\![F(n)]\!]!}}\left(A^{\dagger}\right)^n \vv 0\rr,
\end{equation} 
where $\vv 0\rr=|i\rr$ is the vacuum state of the sector $S_i$ satisfying
$A\vv 0\rr={\cal N}\vv 0\rr=0$ and
$ [\![F(n)]\!]!\equiv F(n)F(n-1)\cdots F(1),\ \ \  [\![F(0)]\!]!
\equiv 1 $. On this basis we have
\begin{equation} 
    A^{\dagger} \vv n\rr = \sqrt{F(n+1)}\vv n+1\rr,\ \ \
    A   \vv n\rr = \sqrt{F(n)}\vv n-1\rr,\ \ \
    {\cal N} \vv n\rr = n \vv n\rr.  \label{reps}
\end{equation} 

\subsubsection{Squeezed vacuum and squeeze operator}

We first consider the squeezed vacuum $|v\rr$ annihilated by $\mu A
+\nu A^{\dagger}$
\begin{equation} 
     \left(\mu A +\nu A^{\dagger} \right)|v\rr = 0,  \label{bv}
\end{equation} 
where the complex numbers $\mu$ and $\nu$ satisfy $|\nu/\mu|<1$.
Let us express it in the form of an exponential displacement-operator
(squeeze operator) acting on the vacuum state.
Expand $|v\rr=\sum_{n=0}^{\infty} C_n \vv n \rr$ and insert it into 
(\ref{bv}), yielding
\begin{equation} 
     C_{2k+1}=0,\ \ \  C_{2k}=C_0 z^k
     \sqrt{\frac{[\![F(2k-1)]\!]!!}{[\![F(2k)]\!]!!}},
\end{equation} 
where $z = -\nu/\mu$, $[\![F(2k)]\!]!!=F(2k)F(2k-2)\cdots F(2)$,
$[\![F(2k-1)]\!]!!= F(2k-1)F(2k-3)\cdots F(1)$ and $[\![F(0)]\!]!!
=[\![F(-1)]\!]!!\equiv 1$.
Then we have
\begin{equation} 
     |v\rr = C_0\sum_{k=0}^{\infty}z^k
             \sqrt{\frac{[\![F(2k-1)]\!]!!}{[\![F(2k)]\!]!!}}\vv 2k\rr 
           = C_0 \sum_{k=0}^{\infty}z^k
             \frac{\left({A^{\dagger 2}}\right)^k}{[\![F(2k)]\!]!!}\vv 0\rr. 
              \label{vv}
\end{equation}  
It is easy to check that the above infinite series converges if $|z|<1$ 
under mild assumptions on the asymptotic behavior of $f(x)$, e.g., 
$f(x)\simeq x^\alpha$ for $x\to\infty$.                      
Now, as a key step, we use the following identity
\begin{equation} 
     \left( \frac{{\cal N}}{F({\cal N})}{A^{\dagger 2}}\right)^k = 
     \left({A^{\dagger 2}}\right)^k
     \frac{{\cal N}+2}{F({\cal N}+2)}\cdots \frac{{\cal N}+2k}{F({\cal N}+2k)}, 
     \label{tt}
\end{equation} 
which, on the vacuum state $\vv 0\rr$, becomes 
\begin{equation} 
     \left( \frac{{\cal N}}{F({\cal N})}{A^{\dagger 2}}\right)^k \vv 0\rr = 
     \left({A^{\dagger 2}}\right)^k
     \frac{[\![(2k)]\!] !!}{[\![F(2k)]\!] !!} \vv 0\rr =
     \left({A^{\dagger 2}}\right)^k
     \frac{2^k k !}{[\![F(2k) ]\!]!!} \vv 0\rr.
\end{equation} 
Then we can rewrite \rref{vv} as
\begin{equation} 
     |v\rr =  C_0
              \sum_{k=0}^{\infty}\frac{1}{k!}
              \left(\frac{z}{2}{A^{\dagger 2}}\right)^k 
              \frac{({\cal N}+2)\cdots ({\cal N}+2k)}
              {F({\cal N}+2)\cdots F({\cal N}+2k)} \vv 0\rr  
           =  C_0 \exp{\left(\frac{z {\cal N}}{2F({\cal N})}{A^{\dagger 2}}\right)}\vv 0\rr.
                        \label{vsol}
\end{equation} 
Following the terminology of the oscillator, the operator
\begin{equation} 
     S(z)=C_0 \exp{\left(\frac{z{\cal N}}{2F({\cal N})}{A^{\dagger 2}} \right)}
\end{equation} 
is referred to as the {\it generalized} squeeze operator.

\subsubsection{Multiphoton coherent states}
 
The multiphoton coherent states are the eigenstate of the 
annihilation operator $A$ 
\begin{equation} 
        A|\alpha\rr = \alpha |\alpha\rr, \label{alpha}
\end{equation} 
where $\alpha$ is an arbitrary complex number. Expanding $|\alpha\rr=
\sum_{n=0}^{\infty} D_n \vv n \rr$ and inserting it into \rref{alpha}, we have
$ D_n=D_0\alpha^n /\sqrt{[\![F(n)]\!]!}$ and 
the coherent state $|\alpha\rr$ is obtained as
\begin{equation} 
       |\alpha\rr = D_0 \sum_{n=0}^{\infty} \frac{\alpha^n}
                    {[\![F(n)]\!]!}
                    (A^{\dagger})^n \vv 0\rr.\label{coherent}
\end{equation} 
Using the following identity
\begin{equation} 
    \left( \frac{{\cal N}}{F({\cal N})} A^{\dagger} \right)^n =
    (A^{\dagger})^n \frac{{\cal N}+1}{F({\cal N}+1)} \cdots  
    \frac{{\cal N}+n}{F({\cal N}+n)} 
    \label{tttt},
\end{equation} 
we obtain the coherent state in the ordinary exponential form
\begin{equation} 
    |\alpha\rr = C_0 \sum_{n=0}^{\infty} \frac{\alpha^n}{n!}
                 \left(\frac{{\cal N}}{F({\cal N})}A^{\dagger} \right)^n \vv 0\rr 
               = C_0 \exp\left(\frac{\alpha{\cal N}}{F({\cal N})}A^{\dagger} \right)
                 \vv 0\rr
               = C_0 D(\alpha)\vv 0\rr.                   \label{cs}
\end{equation} 

We remark that the coherent states of GDO have already been studied
extensively. In particular, Shanta {\it et. al.} have obtained the result
(\ref{cs}) \cite{sha1} using a different method (for the
$q$-oscillator case see \cite{cha2}). They first 
look for an operator $G^{\dagger}$ such that $[A, G^{\dagger}]=1$ and 
then write the displacement operator as $D(\alpha)=\exp{(\alpha 
G^{\dagger})}$.  In fact, by direct verification, we have
          \[ \left[A, \frac{{\cal N}}{F({\cal N})}A^{\dagger} \right]=1. \] 
Our method is using the identity (\ref{tt}) and
(\ref{tttt}), by
which we can easily obtain not only the coherent states but also the
squeezed vacuum states. Furthermore, we shall find that this method
will play an important role in revealing the relevance of the coherent and 
squeezed vacuum states of ISOS to those of the oscillator (see Sec.3.2).
 
In some works the so-called deformed exponential 
displacement operator $\exp_F(\alpha A^{\dagger})\equiv \sum_{n=0}^{\infty}
\frac{\alpha^n(A^{\dagger})^n}{[\![F(n)]\!]!}$ is used to express the coherent state
$\exp_F(\alpha A^{\dagger})\vv 0\rr$. We note that two 
displacement operators $\exp_F(\alpha A^{\dagger})$ and $\exp\left
(\frac{\alpha{\cal N}}{F({\cal N})}A^{\dagger}\right)$ are essentially different, 
although they give rise to the same coherent states by acting on 
the vacuum state.

From the above discussion we see that the $q$-oscillator also 
admits the multi-component squeezed and coherent states through its 
multiphoton realization \rref{q}\thinspace but the relationship is broken 
by the ``vacuum'' expectation value except for in $S_0$. 

\section{GDO and Isospectral oscillator system}
\setcounter{equation}{0}

\subsection{Isospectral oscillator system as a GDO}

In this section we shall first review some basic facts of ISOS and then
show that its creation and annihilation operators and the Hamiltonian
generate a GDO.
The ISOS, as suggested by the name, is a system having the same spectrum
as the ordinary oscillator. The Hamiltonian of the oscillator is 
$H=a^{\dagger}a+\frac{1}{2}=N+\frac{1}{2}$. Then the ISOS Hamiltonian is 
\begin{equation} 
    H_{\lambda}=b^{\dagger}b+\frac{1}{2}=N_{\lambda}+\frac{1}{2},
\end{equation} 
where $b$ and its conjugate $b^{\dagger}$ is defined by
\begin{equation} 
    bb^{\dagger}=aa^{\dagger},
\end{equation} 
and $N_{\lambda}=b^{\dagger}b$. For the realization of the 
operators $b^{\dagger}$ 
and $b$ in the coordinate representation, see \cite{miel,nie1}. In fact,
$\lambda$ is just a parameter entering the coordinate representation
of $b^{\dagger}$ and $b$. From 
the relation
\begin{equation} 
    H_{\lambda}b^{\dagger}=b^{\dagger}(H+1),
\end{equation} 
it follows that the states ($|n-1\rr$ are the eigenstates of $H$)
\begin{equation} 
    |\psi_n\rr=\frac{1}{\sqrt{n}}b^{\dagger}|n-1\rr,\ \ \
    n=1,2,\cdots
\end{equation} 
are the normalized orthogonal eigenstates of $H_{\lambda}$ with
eigenvalues $E_n=n+\frac{1}{2}$. These states, together with the 
state $ |\psi_0\rr$ which is defined by $ b |\psi_0\rr = 0$ and 
is an eigenstate of $H_{\lambda}$ with eigenvalue $1/2$, are complete.
The operators $b^{\dagger}$ and $b$ transform the eigenstates of
$H_{\lambda}$  to those of $H$ and vice versa,
\begin{equation} 
    b^{\dagger}|n\rr= \sqrt{n+1} |\psi_{n+1}\rr,\ \ \ 
    b|\psi_n\rr = \sqrt{n} |n-1\rr.
\end{equation} 
The creation and annihilation operators of ISOS are  found to be
\cite{miel,nie1} 
\begin{equation} 
    A = b^{\dagger} ab, \ \ \ 
    A^{\dagger} = b^{\dagger}a^{\dagger} b.
\end{equation} 
The operators $A$ and $A^{\dagger}$ do not give a closed (Lie) algebra as 
argued in \cite{fern}. Here we are interested in the associative
algebra generated by $A^{\dagger},\ A$ and $N_{\lambda}$ (or $H_{\lambda}$). 
From the above relations, it is not difficult to derive
\begin{equation} 
   [N_{\lambda}, A^{\dagger}]=A^{\dagger},\ \ \ 
   [N_{\lambda}, A  ]=-A, \ \ \
   A^{\dagger} A = \left( N_{\lambda}-1\right)^2 N_{\lambda},\ \ \
   A A^{\dagger} = N_{\lambda}^2 ( N_{\lambda}+1),
\end{equation} 
which is just a GDO with the structure function 
\begin{equation} 
      F(x)=(x-1)^2 x.
\end{equation} 
We denote this algebra by ${\cal B}_{\lambda}$. It is easy to prove
that the algebra ${\cal B}_{\lambda}$ has two orthogonal vacuum 
states $|\psi_0\rr$ and 
$|\psi_1\rr$ defined by $A|\psi_0\rr=A|\psi_1\rr=0$, which correspond
to the two zeroes of $F(x)$. 
The one-dimensional subspace $\{|\psi_0\rr\}$ is invariant and
it forms a one-dimensional representation of ${\cal B}_{\lambda}$
\begin{equation} 
     A|\psi_0\rr = A^{\dagger} |\psi_0\rr = N_{\lambda}|\psi_0\rr = 0,
     \ \ \ \ 
     H_{\lambda}|\psi_0\rr = \frac{1}{2}|\psi_0\rr.
\end{equation} 
The subspace spanned by $\{\ |\psi_n\rr\ |\ n=1,2,\cdots, \ \}$ is 
also an invariant space on which the representation can be easily
obtained as 
\begin{equation} 
    A|\psi_n\rr    = (n-1) \sqrt{n}|\psi_{n-1}\rr, \ \ \
    A^{\dagger} |\psi_n\rr =  n\sqrt{n+1} |\psi_{n+1}\rr, \ \ \
    H_{\lambda}|\psi_n\rr = \left(n+\frac{1}{2}\right)|\psi_n\rr. 
                              \label{ddddd}
\end{equation} 
The representation \rref{ddddd} is an infinite dimensional irreducible 
representation. Therefore the whole Hilbert space is decomposed into
a direct sum of two irreducible subspaces.

\subsection{ Coherent state and squeezed vacuum of ISOS}

We now turn to the coherent and the squeezed vacuum states of the
ISOS, with special emphasis on their relationship with those of the
oscillator. The coherent states of ISOS as the eigenstates of $A$
with the eigenvalue $\alpha$ has already been obtained as \cite{fern}
\begin{equation} 
     |\alpha\rr=\frac{1}{\sqrt{_0 F_2(1,2;|\alpha|^2)}}
           \sum_{n=0}^{\infty}
           \frac{\alpha^n}{n!\sqrt{(n+1)!}} |\psi_{n+1}\rr,
           \label{above}
\end{equation} 
where $ _0 F_2(1,2;|\alpha|^2)$ is a generalized hypergeometric
function defined by \cite{bate}
\begin{equation} 
     _0 F_2(x,y;z)=\sum_{n=0}^{\infty}
     \frac{\Gamma(x)\Gamma(y)}{\Gamma(x+n)\Gamma(y+n)}
     \frac{z^n}{n!}.
\end{equation} 
We now discuss its relationship with some states of the
oscillator, using the {\it identity} techniques presented in Sec.2.3. 
In fact, equation \rref{above} can be rewritten as
\begin{eqnarray} 
     |\alpha\rr &\hspace{-0.3cm}=&\hspace{-0.3cm} b^{\dagger}\frac{1}{\sqrt{\thinspace_0 F_2(1,2,
                        |\alpha|^2)}}\sum_{n=0}^{\infty}\frac{\alpha^n}
                        {n!(n+1)\sqrt{n!}}|n\rr\nonumber \\
                &\hspace{-0.3cm}=&\hspace{-0.3cm} b^{\dagger}\frac{1}{\sqrt{\thinspace_0 F_2(1,2;
                        |\alpha|^2)}}\sum_{n=0}^{\infty}\frac{\alpha^n}
                        {n!(n+1)!}(a^{\dagger})^n |0\rr\nonumber \\ 
                &\hspace{-0.3cm}=&\hspace{-0.3cm} b^{\dagger}\frac{1}{\sqrt{\thinspace_0 F_2(1,2;
                        |\alpha|^2)}}\exp{\left(\frac{\alpha}{N+1}
                        a^{\dagger}\right)}|0\rr.
\end{eqnarray} 
It is easy to see that 
\begin{equation} 
    \left[(N+2)a, \ \frac{1}{N+1}a^{\dagger}\right]=1,
\end{equation} 
therefore, the (unnormalized) state
\begin{equation} 
    \frac{1}{\sqrt{\thinspace_0 F_2(1,2;|\alpha|^2)}}
    \exp{\left(\frac{\alpha}{N+1}a^{\dagger}\right)}|0\rr
\end{equation} 
is the eigenstate of the operator $(N+2)a$, a density-dependent
annihilation operator of the oscillator. Therefore the coherent state can
be obtained by applying the operator $b^{\dagger}$ to the eigenstate 
of the operator $(N+2)a$. 

On the other hand, what is the state obtained by applying the operator $b$ to
$|\alpha\rr$? We in fact have
\begin{equation} 
    b|\alpha\rr = \frac{1}{\sqrt{\thinspace_0 F_2(1,2;|\alpha|^2)}}
             \sum_{n=0}^{\infty}\frac{\alpha^n}{n!n!} 
             (a^{\dagger})^n |0\rr\nonumber \\ 
           = \frac{1}{\sqrt{\thinspace_0 F_2(1,2;|\alpha|^2)}}
             \exp{\left(\frac{\alpha}{N}a^{\dagger}\right)}|0\rr,
\end{equation} 
which is the eigenstate of the density-dependent
annihilation operator $(N+1)a$ of the oscillator. So the coherent
states of the ISO are connected with the eigenstates of the
density-dependent annihilation operators $(N+2)a$ and $(N+1)a$, in
terms of the transformation $b^{\dagger}$ and $b$.

Then we consider the squeezed vacuum defined by 
\begin{equation} 
   \left( \mu A + \nu A^{\dagger} \right)|v\rr=0,\label{ssvv}
\end{equation} 
where the complex numbers $\mu$ and $\nu$ satisfy $|\nu/\mu|<1$.
Taking $|v\rr=\sum_{n=0}^{\infty}C_n |\psi_n\rr$, and inserting it
into the equation \rref{ssvv}, we obtain 
\begin{equation} 
   C_{2k}=0,\ \ \ C_{2k+1}= z^k \left(\frac{(2k-1)!!}{(2k)!!(2k+1)}
                           \right)^{\frac{1}{2}}C_1, \ \ \ \
   z\equiv -\frac{\nu}{\mu}, 
\end{equation} 
Then we have
\begin{equation} 
   |v\rr= C_1 \sum_{k=0}^{\infty}z^k \left(\frac{(2k-1)!!}{(2k)!!(2k+1)}
            \right)^{\frac{1}{2}} |\psi_{2k+1}\rr \nonumber\\
        = b^{\dagger}C_1 \sum_{k=0}^{\infty}\frac{z^k}{(2k+1)k!}
            \left(\frac{a^{\dagger 2}}{2}\right)^{k} |0\rr.
\end{equation} 
The state
\begin{equation} 
   C_1 \sum_{k=0}^{\infty}\frac{1}{2k+1}\frac{z^k}{k!}
            \left(\frac{a^{\dagger 2}}{2}\right)^{k} |0\rr
\end{equation} 
cannot be written in the form of an exponential operator acting on 
the vacuum state $|0\rr$. However, this state can be transformed 
to an exponential state by the action of $b$. It is easy to see that
\begin{equation} 
   b|v\rr = C_1 \sum_{k=0}^{\infty} \frac{1}{k!}\left(z
            \frac{a^{\dagger 2}}{2}\right)^{k} |0\rr.
\end{equation} 
After normalization this state is nothing but the squeezed vacuum of
the oscillator
\begin{equation} 
   b|v\rr = S(z)|\psi_0\rr \equiv \exp\left(z \frac{a^{\dagger 2}}{2}-z^*
            \frac{a^2}{2}\right)|0\rr.
\end{equation} 
Therefore the squeezed vacuum of ISOS is closely related to that of
the oscillator through the transformation $b$.

\section{$q$-GDO and PB phase operator}
\setcounter{equation}{0}

The purpose of this section is twofold: (1) we construct the {\it new} creation
and annihilation operators related with the PB phase operator theory which form
a closed associative algebra (some $q$-deformed GDO) and degenerate to those 
of the
ordinary oscillator in certain limit; (2) we present a
formalism to define algebraically the hermitian phase operator from the
viewpoint of the cyclic representations of some $q$-GDO.

\subsection{$q$-GDO and its cyclic representation}

The GDO {${\cal B}$ } in general does not admit the {\it cyclic} representation.
To connect the GDO with the hermitian phase operator, we have to look for 
the GDO which has {\it finite dimensional cyclic } representations in 
the same sense as in the $q$-oscillator \cite{fuge}. 

Define the algebra {${\cal B}_q$ } as an associative algebra with
generators $A^{\dagger}, A, q^{{\cal N}}$ and 1 subject to the relations
\begin{equation} 
     q^{{\cal N}}A = q^{-1} A \thinspace q^{{\cal N}} , \ \ \ \
     q^{{\cal N}}A^{\dagger} = q A^{\dagger} \thinspace q^{{\cal N}} ,\ \ \ \
     A A^{\dagger}={\cal F}(q^{{\cal N}+1}), \ \ \ \
     A^{\dagger} A = {\cal F}(q^{{\cal N}}), \label{qgdo}
\end{equation} 
where the hermitian non-negative function ${\cal F}$ is again called 
the structure function.
This algebra can be obviously viewed as a subalgebra
of the algebra ${\cal B}$ by identifying ${\cal F}(q^{{\cal N}})\equiv F({\cal N})$.

The algebra ${\cal B}_q$ admits the Fock-like representation for any $q$,
if the condition  ${\cal F}(1)=0$ $\Leftrightarrow$ $F(0)=0$ is 
satisfied.
However, when $q$ is the $(S+1)$-th root of unity, some other types of
representations are possible. Here we are interested in the cyclic
representations, for which the condition ${\cal F}(1)=0$ is not imposed.
 We first prove that, {\it if $q$ is the $(S+1)$-th 
root of unity, the elements $A^{S+1}, (A^{\dagger})^{S+1}$ and 
$(q^{{\cal N}})^{S+1}$ are all the central elements of the algebra {${\cal B}_q$ }.}
This can be shown from (\ref{qgdo}) and the following relations
\begin{eqnarray} 
&&   \left[A^{S+1}, \ A^{\dagger} \right] = A^{S}\left({\cal F}(q^{{\cal N}+1})-
                  {\cal F}(q^{{\cal N}-S})\right)=0, \nonumber \\
&&   \left[A, \ (A^{\dagger})^{S+1}\right] = \left({\cal F}(q^{{\cal N}+1})-
                  {\cal F}(q^{{\cal N}-S})\right) (A^{\dagger})^{S}=0,
\end{eqnarray} 
since $q^{{\cal N}+1}=q^{{\cal N}-S+(S+1)}=q^{{\cal N}-S}$.

Now let us construct the explicit cyclic representation of $q$-GDO when 
$q^{S+1}=1$. Let ${\cal H}_{S}$ be a vector space with an orthonormal basis
\begin{equation} 
    {\cal H}_{S+1}:\ \left\{\  |k\rr \ |\ k=0, 1, 2, \cdots, S \ \right\}.
\end{equation} 
Define the action of ${\cal B}_q$ on ${\cal H}_{S}$ as
\begin{eqnarray} 
&&   A|k\rr      =   \sqrt{{\cal F}(q^{k+\eta})} |k-1\rr,\ \ \ 
                       k\neq 0, \ \ \ \ \ \ \ \ 
     A|0\rr      =   \xi^{-1}\sqrt{{\cal F}(q^{\eta})}|S\rr,
                       \ \ \ \xi\neq 0, \nonumber\\
&&   A^{\dagger} |k\rr   =   \sqrt{{\cal F}(q^{k+\eta+1})} |k+1\rr,\ \ \ 
                       k\neq S, \ \ \ \
     A^{\dagger} |S\rr   =   \xi\sqrt{{\cal F}(q^{\eta})} |0\rr, \nonumber \\
&&   q^{{\cal N}} |k\rr  =   q^{k+\eta} |k\rr, \label{cyc}
\end{eqnarray} 
where $\eta$ is an extra  real parameter (which may depend on $S$) 
\cite{OKK}
and $\xi\neq 0$ is a complex constant. 
One can directly verify that Eqs.\rref{cyc} define an ($S$+1)-dimensional
cyclic representation of {${\cal B}_q$ } if 
\begin{equation} 
       {\cal F}(q^{\eta+k}) \neq 0 \quad {\rm for}\quad k=0,1, \cdots,S.
       \label{novancond}
\end{equation}    
In this representation the central elements take
\begin{eqnarray} 
&\hspace{-0.3cm}&\hspace{-0.3cm}   (q^{{\cal N}})^{S+1} = q^{\eta (S+1)}, \ \ \
       (A^{\dagger})^{S+1} = \xi\sqrt{{\cal F}(q^{\eta}){\cal F}(q^{\eta+1})\cdots
                     {\cal F}(q^{\eta+S})}, \nonumber \\
&\hspace{-0.3cm}&\hspace{-0.3cm}   A^{S+1}     
        = \xi^{-1}\sqrt{{\cal F}(q^{\eta}){\cal F}(q^{\eta+1})\cdots
                     {\cal F}(q^{\eta+S})},
\end{eqnarray} 
which are non-vanishing constants.

It should be remarked that the naive  $q$-oscillator with 
$q^{S+1}=1$, namely
\begin{equation} 
       {\cal F}(q^{{\cal N}})= [{\cal N}]= 
       \left(q^{{\cal N}}-q^{-{\cal N}}\right)/(q-q^{-1})
        \label{qosciF}
\end{equation} 
{\it fails to provide} a cyclic representation simply because 
${\cal F}(q^{{\cal N}})$ takes negative as well as positive values. 
Some admissible choices are:
\begin{eqnarray} 
&\hspace{-0.3cm}&\hspace{-0.3cm}   {\cal F}(q^{{\cal N}})= |[{\cal N}]|=
       \left|\left(q^{{\cal N}}-q^{-{\cal N}}\right)/(q-q^{-1})\right|,
       \label{absN}\\
&\hspace{-0.3cm}&\hspace{-0.3cm}   {\cal F}(q^{{\cal N}})=  
       \left|\left(q^{{\cal N}}-q^{-{\cal N}}\right)/(q-q^{-1})+K(S)\right|,
       \label{conAdd}
\end{eqnarray} 
where $K(S)$ is real. Let us call them positive `$q$-oscillators'.
For the case (\ref{absN}), the condition $|[k+\eta]|\neq0$ is 
satisfied if $\eta$ is not an integer. 
Then (\ref{cyc}) defines a cyclic representation.
For the case of (\ref{conAdd}) we can also choose $\eta$ and $K(S)$ to 
have cyclic representations by satisfying (\ref {novancond}). 
For example,
\begin{equation} 
0<|K(S)|<{1\over2}\quad {\rm and}\quad \eta\equiv0.
\end{equation} 

We will return to these examples in connection with the hermitian 
phase operator.

Another well known type of $q$-deformed oscillator ($q$ complex)
\begin{equation} 
        {\cal F}(q^{{\cal N}})=\frac{1-q^{{\cal N}}}{1-q}
\end{equation} 
or equivalently,
\begin{equation} 
        A A^{\dagger}  - qA^{\dagger} A= 1
\end{equation} 
is not admissible as a $q$-GDO. It is obvious that the
{\it hermiticity condition} of ${\cal F}$ is not satisfied.

Another example of the $q$-GDO is the dynamical symmetry algebra of 
the hamiltonian system with self-similar potentials \cite{self}. 
In this system the  symmetry algebra is ($q$: real)
\begin{equation} 
    L B^{\dagger}= q^2 B^{\dagger}L, \ \ \ \
    L B = q^{-2} BL, \ \ \ \
    B^{\dagger}B= \prod_{n=0}^{M}(L+\omega_n),\ \ \ \
    BB^{\dagger}= \prod_{n=0}^{M}(q^2 L+\omega_n), \label{self}
\end{equation} 
where $L$ is the Hamiltonian, $\omega_n$ are some real and positive
constants  and $M$
is a positive integer. It can be rewritten in the form (\ref{qgdo}) by 
identification 
\begin{equation} 
   L \longrightarrow \left(q^{{\cal N}}\right)^2,\ \ \ \
   B^{\dagger} \longrightarrow A^{\dagger}, \ \ \ \
   B \longrightarrow A.
\end{equation} 
 
Next we shall prove that the $q$-GDO can also be realized 
in terms of the PB phase operators in quantum optics and on the other
hand it can be used to define the hermitian phase operator.

\subsection{Creation and annihilation operators of PB 
            phase operator theory}

Let us begin with PB's theory of hermitian phase operator.  The PB phase 
operator is defined in an $(S+1)$-dimensional space ${\cal H}_S$ 
spanned by the {\it number} basis $|n\rr, \ n=0,1,\cdots,S$, with the
inner product $\langle m |n \rangle = \delta_{mn}$.  Define the 
{\it phase states} $|\theta_m\rr,\ m=0,1,\cdots,S$,
\begin{equation} 
    |\theta_m\rr = \frac{1}{\sqrt{S+1}}\sum_{n=0}^S \exp(in\theta_m)|n\rr,
\end{equation} 
where $\theta_m = \theta_0+\frac{2\pi m }{S+1}$ and $\theta_0$ are real
constants. Hereafter we write $\exp(in\theta_m)$ as
\begin{equation} 
    \exp(in\theta_m)=\exp(in\theta_0)q^{mn}, \label{thetata}
\end{equation} 
where $q$ is the deformation parameter
\begin{equation} 
    q=\exp\left(\frac{2\pi}{S+1}i\right)
\end{equation} 
satisfying $q^{S+1}=1$. From the orthonormality of the number states 
$\langle m|n\rr=\delta_{mn}$ it is
easy to prove that of the phase states 
$\langle \theta_m|\theta_n\rr=\delta_{mn}$. 
We can express the number states in terms of the phase states
\begin{equation} 
      |n\rr=\frac{1}{\sqrt{S+1}}\sum_{m=0}^{S} \exp{(-in\theta_m)}
      |\theta_m\rr.
\end{equation} 
The {\it PB phase operator} is defined as
\begin{equation} 
      \hat{\Phi}_{\theta}=\sum_{m=0}^{S} \theta_m |\theta_m \rr \langle
      \theta_m |;  \ \ \ \ \
      \hat{\Phi}_{\theta} |\theta_m\rr= \theta_m |\theta_m\rr.
\end{equation} 
A representation of the exponential PB phase operator
$\p$ on $|n\rr$ is obtained as
\begin{eqnarray} 
&&     \p   |n\rr  =  |n-1\rr ,   \ \  \ \ n\neq 0, \nonumber \\
&&     \p   |0\rr  =  \exp{\{i(S+1)\theta_0\}}|S\rr,  \nonumber\\
&&     \mbox{e}^{-i\hat{\Phi}_{\theta}}  |n\rr  =  |n+1\rr, \ \ \ \ \ n \neq S,  \nonumber\\
&&     \mbox{e}^{-i\hat{\Phi}_{\theta}}  |S\rr  =  \exp{\{-i(S+1)\theta_0\}}|0\rr. \label{cycc}
\end{eqnarray} 
At this stage, Pegg and Barnett defined the creation and the
annihilation operators
\begin{equation} 
       a_{\mbox{\tiny PB}}^{\dagger}=\sqrt{\hat N}\mbox{e}^{-i\hat{\Phi}_{\theta}}, \ \ \ \
        a_{\mbox{\tiny PB}}=\p \sqrt{\hat N},
\end{equation} 
where 
\begin{equation} 
       {\hat N}= \sum_{n=0}^{S} n|n\rr \langle n |.
\end{equation} 
Then $a_{\mbox{\tiny PB}}^{\dagger}$ and $a_{\mbox{\tiny PB}}$
satisfy the so-called {\it truncated oscillator} commutation relation
\begin{equation} 
        [a_{\mbox{\tiny PB}}, a_{\mbox{\tiny PB}}^{\dagger}]=
        1-(S+1)|S\rr \langle S |,
\end{equation} 
which they claim to degenerate to that of the ordinary  oscillator
\begin{equation} 
        \langle p |[a_{\mbox{\tiny PB}}, a_{\mbox{\tiny
        PB}}^{\dagger}]|p\rr_{S\rightarrow \infty} =1,
\end{equation} 
on the ``physical states'' $|p\rr$ (for example, on the coherent
states of the single mode electromagnetic field) in
the limit $S\rightarrow \infty$.  We note that the truncated 
oscillator does not form a closed algebra and that
the operator relations do not simply reduce to those of the ordinary
oscillator  in the limit $S\rightarrow \infty$.

Here we shall define  new creation and annihilation operators, which
form a {\it closed algebra} (some $q$-GDO introduced in
Sec.4.1) and degenerate to the usual oscillator in the limit $S\to
\infty$.  For this purpose, let us define
\begin{equation} 
     A^{\dagger}=\sqrt{{\cal F}(q^{{\cal N}})}\ 
                 e^{-i\hat{\Phi}_{\theta}},\ \ \ \ 
     A =\p \sqrt{{\cal F}(q^{{\cal N}})},   \ \ \ \
     q^{{\cal N}}= q^{{\hat N}+\eta}= e^{i\frac{2\pi \eta}{S+1}}
              e^{i\frac{2\pi}{S+1}{\hat N}},\label{def}
\end{equation}  
where the operator $e^{i\frac{2\pi}{S+1}{\hat N}}$ is nothing but the
{\it phase shift operator}
\begin{equation} 
      e^{i\frac{2\pi}{S+1}{\hat N}}|\theta_m\rr=
      \left| \theta_m + \frac{2\pi}{S+1}\right\rr \equiv
      |\theta_{m+1}\rr, \label{shift}
\end{equation} 
and the parameter $\eta$ and the function ${\cal F}$ will be
specified later. It is easy to see that Eqs.(\ref{def})
lead to the following relations
\begin{eqnarray} 
&&    A^{\dagger} |n\rr  = \sqrt{{\cal F}(q^{n+\eta +1})}\: |n+1\rr,\ \ \  n\neq
                     S,\nonumber \\
&&    A^{\dagger} |S\rr  = \exp{\{-i(S+1)\theta_0\}} \sqrt{{\cal F}(q^\eta)}\,
                     |0\rr,  \nonumber \\
&&    A   |n\rr  = \sqrt{{\cal F}(q^{n+\eta})}\, |n-1\rr,\ \ \ n\neq 0,
                     \nonumber \\
&&    A   |0\rr  = \sqrt{{\cal F}(q^\eta)} \exp{\{i(S+1)\theta_0\}}\,
                     |S\rr,  \nonumber \\
&&    q^{{\cal N}}  |n\rr  = q^{n+\eta}|n\rr, \label{cyccc}
\end{eqnarray} 
which are just the representation (\ref{cyc}) of the $q$-GDO
with $\xi=\exp{\{-i(S+1)\theta_0\}}$. Therefore the operators defined
in (\ref{def}) generate a $q$-GDO. 

Now we consider the constraints on the parameter $\eta$ or $K(S)$ 
in the function ${\cal F}$. 
The following conditions should be satisfied:
\begin{enumerate}
\item    The function ${\cal F}$ must be hermitian and non-negative due to the 
same properties of the operator $A^\dagger A$.
\item     Choose $\eta$ such that the representation (\ref{cyccc}) is a cyclic 
representation.
\item
In the classical limit $S\to \infty$ ($q\to 1$) or zero deformation, 
the operators $A^\dagger$, $A$ 
should tend to the  creation and annihilation operators of the 
ordinary oscillator.
\end{enumerate}
The condition for the cyclic representation is equivalent to
the condition that the operator ${\cal F}(q^{\cal N})$ has the inverse 
which is necessary in order to define the PB phase operator. 
For simplicity and concreteness, let us discuss the positive `$q$-oscillators'
(\ref{absN}),(\ref{conAdd}). 
As mentioned above 
the condition for non-vanishing central elements (\ref{novancond}) can be easily 
satisfied for (\ref{absN}) case if $\eta(S)$ is not an integer.
As for the other example  (\ref{conAdd}), the same condition is 
satisfied for example by
\begin{equation} 
     \eta\equiv0,\quad 0<|K(S)|<{1\over2}.
     \label{Kcond}
\end{equation} 
In either case, the algebra of $A$ and $A^\dagger$ degenerate to 
that of the ordinary oscillator (Weyl algebra)  provided
\begin{equation} 
      \lim_{S\to\infty}\eta(S)=0,\quad {\rm or} \qquad      
      \lim_{S\to\infty}K(S)=0.
      \label{redcond}
\end{equation}    
These can be achieved, for example, $\eta(S)=K(S)={1\over{S+1}}$.   

We would like to remark that Ellinas \cite{ell2} studied the phase
operator from the {\it regular} representation of the naive 
$q$-oscillator with $q^{S+1}=1$. As remarked in the previous 
subsection, the naive $q$-oscillator (\ref{qosciF}) with $q$ 
root of unity does not possess an admissible algebraic structure 
to connect with the hermitian phase operator. Moreover, in the 
regular representation characterized by the condition
$A|0\rangle =A^\dagger|S\rangle=0$, both of the operators $A$ and 
$A^\dagger$ have a zero mode. Therefore the polar decomposition 
for them does not exist and the hermiticity of the phase operator, 
if any, is not guaranteed.

\subsection{Algebraic definition of the hermitian phase operator}

Now we turn to the study of the hermitian phase operator from the viewpoint of 
the cyclic representations of $q$-GDO. For concreteness, we consider the
positive `$q$-oscillator' (\ref{absN}) only. For the other choices of ${\cal 
F}$ like (\ref{conAdd}) the discussion is essentially the same.
 We start with the cyclic
representation (\ref{cyc}) of positive `$q$-oscillator'  on the space ${\cal
H}_S$ with the inner product $\langle m | n \rangle = \delta_{mn}$.
Define the exponential phase operator $\p$ and 
$\mbox{e}^{-i\hat{\Phi}_{\theta}}$ by the 
relation (\ref{def}). Since we have chosen $\eta$ such that
$\sqrt{|[\eta+k]|}\neq 0$ ($k\in \{0,1,\cdots,S\}$) in
the cyclic representation (\ref{def}), the operator $\sqrt{|[{\cal N}]|}$ 
 (\ref{cyc}) has the inverse.
Therefore in the cyclic representation (\ref{cyc}) the operators
$e^{\pm i{\hat \Phi}_\theta}$ are  well defined uniquely:
\begin{equation} 
   e^{i{\hat \Phi}_\theta} = \{|[{\cal N}+1]|\}^{-\frac{1}{2}} A,\ \ \ \
   e^{-i{\hat \Phi}_\theta} = \{|[{\cal N}]|\}^{-\frac{1}{2}}A^{\dagger}.
     \label{ddeeff}
\end{equation} 
Inserting (\ref{ddeeff}) into (\ref{cyc}), we find the action of 
$e^{\pm i{\hat \Phi}_\theta}$ on ${\cal H}_S$
\begin{eqnarray} 
&&  e^{i{\hat \Phi}_\theta}  |k\rr = |k-1\rr, \ \ \ \ k \neq 0,  \ \ \ \
    e^{i{\hat \Phi}_\theta}  |0\rr = \xi^{-1} |S\rr,  \nonumber\\
&&  e^{-i{\hat \Phi}_\theta} |k\rr = |k+1\rr,\ \ \ \ k \neq S, \ \ \ \
    e^{-i{\hat \Phi}_\theta} |S\rr = \xi |0\rr, \nonumber \\
&&  q^{\cal N}             |k\rr = q^{k+\eta} |k\rr.\label{4155}
\end{eqnarray} 
Choosing $\xi=e^{-i\theta_0 (S+1)}$, we exactly reproduce PB's exponential
phase operator (\ref{cyccc}).

It is  not convenient to derive the action of phase operator
$\hat{\Phi}_{\theta}$ itself on the number states. To evaluate the
phase operator $\hat{\Phi}_{\theta}$ itself, we have to look for a
basis on which $\p$ is diagonal. 
To this end we evaluate the eigenstates of $\p$
\begin{equation} 
    \p |z\rr=z|z\rr. \label{41}
\end{equation} 
Suppose that $|z\rr=\sum_{n=0}^{S}C_n|n\rr$, where $C_n$ is coefficient
to be determined. Then inserting it into the Eqs.\rref{41} we obtain $S+1$
distinct eigenvalues
\begin{equation} 
    z_m=\exp(i\theta_0)\exp\left(\frac{2\pi m i}{S+1}\right)\equiv
    \exp(i\theta_m), \ \ \ m=0,1,\cdots,S,
\end{equation} 
where $\theta_m$ is same as in Eq.(\ref{thetata}). Then the corresponding
eigenstates are
\begin{equation} 
    |\theta_m\rr \equiv |z_m\rr = C_0 \sum_{n=0}^{S}\exp(i\theta_m n) |n\rr,
\end{equation} 
and their inner product
\begin{equation} 
    \langle \theta_m | \theta_n \rangle = |C_0|^2 (S+1) \delta_{mn}.
\end{equation} 
Requiring that the states $|\theta_m \rangle$ are normalized, the
constant $C_0$ is fixed as $\frac{1}{\sqrt{S+1}}$. 
In comparison with PB's theory, these eigenstates form the phase states.
On the phase states the eigenvalue equation (\ref{41}) becomes
\begin{equation} 
    \p |\theta_m\rr = \exp(i\theta_m) |\theta_m\rr,
\end{equation} 
from which we can {\it define} the hermitian phase operator 
$\hat{\Phi}_{\theta}$ as follows
\begin{equation} 
     \hat{\Phi}_{\theta} |\theta_m\rr = \theta_m |\theta_m\rr.
\end{equation} 
So, this approach exactly recovers PB's theory.
It should be noted, however, that all the eigenvalues could be shifted  
by an integer multiple of $2\pi$, which is natural for 
a phase. In other words it can be absorbed by the redefinition of 
$\theta_0$.

\section{Conclusion}
\setcounter{equation}{0}

In this paper we have studied the GDO and some of its properties, namely,
the multiphoton realizations, the ladder-operator coherent and 
squeezed vacuum states. 
The coherent displacement-operator $D(\alpha)$ 
and the squeeze operator $S(z)$ are explicitly constructed and
expressed in the exponential form. For the ordinary oscillator, 
we know that the 
state $D(\alpha)S(z)|0\rr$ 
\begin{equation} 
      |\alpha,z\rr \equiv D(\alpha)S(z)|0\rr=
                  C_0 e^{\alpha a^{\dagger}}e^{z a^{\dagger 2}}|0\rr 
                  \stackrel{\mbox{\footnotesize normalization}}
                    {=\!=\!=\!=\!=\!=\!=\!=\!=}
                    e^{\alpha a^{\dagger}-\alpha^* a}
                    e^{z a^{\dagger 2}-z^* a^2}|0\rr,    
\end{equation} 
is just the squeezed state, which is also 
the eigenstate of $\mu a + \nu a^{\dagger}$. However, for the general
case (the GDO), the state $D(\alpha)S(z)\vv 0\rr$ is not the squeezed state
equivalent to the ladder-operator definition, namely, it is
not the eigenstate of $\mu A +\nu A^{\dagger}$. The coherent
displacement-operator $D(\alpha)$ is a {\it good} operator in the
sense that it enjoys the following property
\begin{equation} 
      D(-\alpha)AD(\alpha)=A+\alpha.
\end{equation}  
However, the squeeze operator $S(z)$ does not keep the Bogoliubov
transformation, namely
\begin{equation} 
     S^{-1}(z)AS(z)\neq \mu A + \nu A^{\dagger}.
\end{equation} 
This is why the state $D(\alpha)S(z)|0\rr$ is not the eigenstate of
$\mu A+\nu A^{\dagger}$, as is argued in \cite{nie2}. 
However, we can expect that the states $D(\alpha)S(z)\vv 0\rr$ and 
$S(z)D(\alpha)\vv 0\rr$ are
important quantum states in quantum optics and it is a good challenge
to study their nonclassical properties.

We have pointed out that a realistic physical system, the ISOS 
(isospectral oscillator system), has GDO as
its symmetry algebra. Its coherent and squeezed vacuum 
states are studied in some detail and  that they are compared and  related 
with  those of
the ordinary oscillator.

To connect the GDO with the hermitian phase operator, we have introduced a
new algebra, the $q$-GDO, which is a subalgebra of GDO depending on a
complex parameter $q$.  It has  cyclic representations when $q$ is a root
of unity. 
This approach has two remarkable advantages:
(1) The phase operator of the Pegg-Barnett's theory can be 
constructed from the $q$-GDO {\it purely algebraically\/}; 
(2) The $q$-GDO with
$q^{S+1}=1$ provides a finite dimensional space to define the phase
operator and the cyclic representations ensure the hermiticity of the
phase operator in contrast with the regular representation case.  


\section*{Acknowledgments}

We thank D.\thinspace Fairlie, P.\thinspace Kulish, M.\thinspace Pillin and C.\thinspace Zachos 
for useful comments. H.\thinspace C.\thinspace Fu is grateful to Japan Society for 
Promotion of Science (JSPS) for the fellowship.
He is also supported in part by the National Science Foundation of
China.

\section*{Appendix A: Multimode GDO}
\setcounter{equation}{0}
\renewcommand{\theequation}{A.\arabic{equation}}

The formalism in Sec.2.1 can be easily generalized to the multimode 
case. For simplicity we consider only the two-mode case. 
Generalization to three-mode and further is straightforward.
Consider the two-mode photon field described by two independent modes
\begin{equation} 
    [a, a^{\dagger}]=1,\ \ [b, b^{\dagger}]=1,
\end{equation} 
and an arbitrary two-mode multiphoton annihilation oscillator
\begin{equation} 
    A=f(N_1,N_2)a^m b^n,
\end{equation} 
where $N_1=a^{\dagger}a,\ N_2=b^{\dagger}b$, $f$ is an arbitrary
function with $f(n_1,n_2)\neq 0$ for $ n_1,n_2$ non-negative
integers. Note that $f$ is not necessarily factorized as $f(N_1,N_2)
=f_1(N_1)f_2(N_2)$. It is easy to see $(i=1,2)$
\begin{eqnarray} 
    AA^{\dagger}=F({\cal N}_1+1,{\cal N}_2+1), \ \ \ 
       A^{\dagger} A= F({\cal N}_1,{\cal N}_2), \ \ \
    [{\cal N}_i, A^{\dagger}]=A^{\dagger},\ \ \ [{\cal N}_i, A]=-A, 
\end{eqnarray} 
where
\begin{eqnarray} 
&&    {\cal N}_1\equiv \frac{1}{m}(N_1-i), \ \ \ \
      {\cal N}_2\equiv \frac{1}{n}(N_1-j), \ \ \ \ 
                  (0\leq i\leq m-1,\ 0\leq j\leq n-1),  \nonumber \\
&&    F({\cal N}_1+1,{\cal N}_2+1) \equiv (N_1+1)\cdots(N_1+m)
                  (N_2+1)\cdots(N_2+n) f(N_1,N_2)f^*(N_1,N_2)\nonumber \\
&&    \ \ \ \ \ \ \equiv (m{\cal N}_1+i+1)\cdots (m{\cal N}_1+i+m)(n{\cal N}_2+j+1)
                  \cdots  \nonumber \\
&&    \ \ \ \ \ \ \ \ \  (n{\cal N}_2+j+n) f({\cal N}_1,{\cal N}_2)f^*({\cal N}_1,{\cal N}_2).
\end{eqnarray} 
This algebra is defined in a subspace $\bar{S}_{ij}$ of the 
sector $S_{ij}$ spanned by ($ k=0,1,2,\cdots $)
\begin{equation} 
       \vv k\rr \equiv \frac{1}{\sqrt{F(k,k)!}}
                    (A^{\dagger})^k |i,j\rr \propto |km+i,kn+j\rr.
\end{equation} 
The representation on $\bar{S}_{ij}$ is
\begin{equation} 
   A^{\dagger} \vv k\rr = \sqrt{F(k,k)}\vv k+1\rr, \ \ \ \
   A   \vv k\rr = \sqrt{F(k,k)}\vv k-1\rr, \ \ \ \
   {\cal N}_1 \vv k\rr = {\cal N}_2 \vv k\rr = k \vv k\rr.  \label{rree}
\end{equation} 

We consider the eigenvalue equation
\begin{equation} 
     \left(\mu A+\nu A^{\dagger}\right)|\beta\rr=\beta|\beta\rr.
\end{equation}  
These states are degenerate. The degeneracy can be lifted by assuming
that the $(m+n)$ photons are either created or annihilated
together. This means the following conservation law
\begin{equation} 
     \left({\cal N}_1 - {\cal N}_2\right)|\beta\rr = 0.
\end{equation} 
In the representation \rref{rree} the condition is fulfilled
automatically.

Identifying $F(k,k)$ here with $F(k)$ in section 2, the representation
\rref{rree} takes the same form as \rref{reps}. So, formally, the squeezed 
states can be investigated in the same manner as those in section 2.2.


\end{document}